\documentclass[twocolumn]{aastex631}

\usepackage{verbatim}
\usepackage{graphicx}

\usepackage{savesym}
\savesymbol{tablenum}
\usepackage{amsmath,amssymb,siunitx}

\received{\today}
\revised{\today}

\submitjournal{RNAAS}

\shorttitle{FRB Scintillation Timescales}

\shortauthors{Schoen et al.}
\graphicspath{{./}{figures/}}

\begin{document}


\title{Scintillation Timescales of Bright FRBs Detected by CHIME/FRB}

\correspondingauthor{Eve Schoen}
\email{eveschn@mit.edu}

\author[0000-0001-8360-2098]{Eve Schoen} 
\affiliation{MIT Kavli Institute for Astrophysics and Space Research, Massachusetts Institute of Technology, 77 Massachusetts Ave, Cambridge, MA 02139, USA}
\author[0000-0002-4209-7408]{Calvin Leung}
\affiliation{MIT Kavli Institute for Astrophysics and Space Research, Massachusetts Institute of Technology, 77 Massachusetts Ave, Cambridge, MA 02139, USA}
\affiliation{Department of Physics, Massachusetts Institute of Technology, 77 Massachusetts Ave, Cambridge, MA 02139, USA}
\author[0000-0002-4279-6946]{Kiyoshi Masui}
\affiliation{MIT Kavli Institute for Astrophysics and Space Research, Massachusetts Institute of Technology, 77 Massachusetts Ave, Cambridge, MA 02139, USA}
\affiliation{Department of Physics, Massachusetts Institute of Technology, 77 Massachusetts Ave, Cambridge, MA 02139, USA}
\author[0000-0002-2551-7554]{Daniele Michilli}
\affiliation{MIT Kavli Institute for Astrophysics and Space Research, Massachusetts Institute of Technology, 77 Massachusetts Ave, Cambridge, MA 02139, USA}
\affiliation{Department of Physics, Massachusetts Institute of Technology, 77 Massachusetts Ave, Cambridge, MA 02139, USA}

\author[0000-0002-3426-7606]{Pragya Chawla}
\affiliation{Department of Physics, McGill University, 3600 rue University, Montr\'eal, QC H3A 2T8, Canada}
\affiliation{McGill Space Institute, McGill University, 3550 rue University, Montr\'eal, QC H3A 2A7, Canada}
\affiliation{Anton Pannekoek Institute for Astronomy, University of Amsterdam, Science Park 904, 1098 XH Amsterdam, The Netherlands}

\author[0000-0002-8912-0732]{Aaron B.~Pearlman}
\altaffiliation{McGill Space Institute~(MSI) Fellow.}
\altaffiliation{\mbox{\hspace{-0.29cm}}$^{\ast}$ FRQNT Postdoctoral Fellow.}
\affiliation{Department of Physics, McGill University, 3600 rue University, Montr\'eal, QC H3A 2T8, Canada}
\affiliation{McGill Space Institute, McGill University, 3550 rue University, Montr\'eal, QC H3A 2A7, Canada}
\affiliation{Division of Physics, Mathematics, and Astronomy, California Institute of Technology, Pasadena, CA 91125, USA}
\author[0000-0002-6823-2073]{Kaitlyn Shin}
\affiliation{MIT Kavli Institute for Astrophysics and Space Research, Massachusetts Institute of Technology, 77 Massachusetts Ave, Cambridge, MA 02139, USA}
\affiliation{Department of Physics, Massachusetts Institute of Technology, 77 Massachusetts Ave, Cambridge, MA 02139, USA}
\author[0000-0001-5351-824X]{Ashley Stock}
\affiliation{David A.~Dunlap Department of Astronomy \& Astrophysics, University of Toronto, 50 St.~George Street, Toronto, ON M5S 3H4, Canada}
\affiliation{Canadian Institute for Theoretical Astrophysics, University of Toronto, 60 St.~George Street, Toronto, ON M5S 3H8, Canada}

\collaboration{10}{(CHIME/FRB Collaboration)}



\begin{abstract}
We describe a pipeline to measure scintillation in fast radio bursts (FRBs) detected by CHIME/FRB in the 400-800 MHz band by analyzing the frequency structure of the FRB's spectrum. We use the pipeline to measure the characteristic frequency bandwidths of scintillation between $4-\SI{100}{\kilo\hertz}$ in 12 FRBs corresponding to timescales of $\sim$2-40 $\mu$s for 10 FRBs detected by CHIME/FRB. For the other two FRBs, we did not detect scintillation in the region our analysis is sensitive. We compared the measured scintillation timescales to the NE2001 predictions for the scintillation timescales from the Milky Way. We find a strong correlation to be an indication that in most instances, the observed scintillation of FRBs can be explained by the Milky Way. 
\end{abstract}



\section{Introduction}
\label{sec:intro}
FRBs are a new class of extragalactic millisecond-duration radio transients~\citep{lorimer2007bright, Petroff_2019}. The unique design of the Canadian Hydrogen Intensity Mapping Experiment (CHIME) has enabled CHIME/FRB \citep{FRBSystemOverview} to advance the study of FRBs as a population~\citep{CHIMEFRB_CAT1}. However, many key aspects of FRBs, including their host galaxies and progenitors, remain unclear. 

Plasma lensing is a potent probe of the FRB's host galaxy, the Milky Way, and even the host environment of the FRB. Radio waves stochastically interfere as they propagate through inhomogeneities in a plasma, introducing a characteristic plasma lensing timescale into radio emission. Plasma lensing is readily detectable in fast radio transients, e.g. pulsars and FRBs, and the measured timescale is sensitive to the lensing geometry and plasma properties. Depending on whether the characteristic timescale is longer or shorter than the instrumental time resolution ($\approx \SI{100}{\micro\second}$ for CHIME/FRB), plasma lensing is observable in radio transients as either temporal pulse broadening (``scattering'') with timescale $\tau$  or as banded structures in the spectrum of the signal (``scintillation'') on a characteristic scale called the decorrelation bandwidth, hereafter called $\nu_{dc}$. In this work we present measurements of decorrelation bandwidths from a sample of bright FRBs using baseband data collected by CHIME/FRB ~\citep{michilli2020analysis}. 

\section{Selection Criteria}
We selected 15 bright FRBs, three of which are taken from the CHIME/FRB catalog, with the criteria that their signal-to-noise ratio as measured in a dynamic spectrum by \texttt{fitburst}, a CHIME/FRB burst fitting script, was above 70~\citep{CHIMEFRB_CAT1}. We included \textbf{9} bursts from apparently non-repeating FRBs and \textbf{3} bursts from FRBs sources observed to repeat. Of the bursts in our sample, 4 have scattering tails reported to be shorter than $\SI{100}{\micro\second}$---the shortest scattering tail to which \texttt{fitburst} is sensitive.

\section{Scintillation Pipeline}
To measure the short ($\lesssim \SI{100}{\micro\second}$) characteristic timescales in FRBs, it is possible to measure $\nu_{dc}$ by studying the fine-scale structure of the FRB's spectrum. Our pipeline uses beamformed voltage data from the CHIME/FRB baseband system~\citep{michilli2020analysis}. It rechannelizes baseband data to high spectral resolution ($\SI{24}{\kilo\hertz}$), calculates a spectrum, autocorrelates them, and fits a function to find the decorrelation bandwidth. 
To do this, we coherently de-dispersed each FRB to the dispersion measure (DM) that maximizes the height of the pulse peak when summed over the 400 MHz bandwidth. Next, the time window region where the FRB is present (the on-region henceforth) is selected. Frequency channels with radio frequency interference (RFI) are masked so that they do not contaminate the spectral structure of the FRB. Finally, we decreased the time resolution and increased the frequency resolution of the spectra by a factor of 16 (to \SI{24}{\kilo\hertz} resolution) by Fourier transforming each frequency channel along the time axis. We square the voltage data to obtain flux units, and then integrate over the time duration of the FRB. We also sum both telescope polarizations to suppress the contaminating effect of Faraday rotation on our signal. 

To characterize statistical noise in our data, this process is repeated for regions of noise before and after the FRB signal (off-pulse regions) of the same duration as the FRB. We estimate the system temperature of the telescope by subtracting the mean of $\sim 50$ off-pulse spectra ($\langle S_{\rm off}(\nu)\rangle$) from the on-pulse spectrum ($S_{\rm on}(\nu)$). In addition, we perform a second round of RFI cleaning by removing 3$\sigma$ outliers in the average of all off-pulse spectra.

Then we fit a spline to ($S_{\rm on}(\nu) - \langle S_{\rm off}(\nu)\rangle$) to create a smoothed spectrum, $\bar{S}(\nu)$. We divide by the smoothed spectrum to remove slowly-varying ($\gtrsim \SI{7}{\mega\hertz}$) frequency structures present in the FRB and spectral structures from the telescope's primary beam. The spline has knots separated by 7 MHz, restricting the parameter space to scintillation bandwidths significantly smaller than that.  Nevertheless, fine frequency structure due to $\gtrsim$ microsecond-scale Galactic scintillation is still present in the combination $\delta_{\rm on}(\nu) = \dfrac{S_{\rm on}(\nu) - \langle S_{\rm off}(\nu) \rangle}{\bar{S}(\nu)}-1$.
To detect scintillation we calculate the frequency autocorrelation functions (ACFs) of $\delta_{\rm on}(\nu)$ at high spectral resolution over CHIME's operating range of 400-800 MHz. We define the ACF at frequency channel lag $\Delta \nu$ as $\textbf{r}(\Delta \nu)$ where:
\begin{equation}
    \mathbf{r}(\Delta \nu) = \dfrac{1}{N} \sum_{\nu} \delta(\nu) \delta(\nu + \Delta \nu).
    \label{eq:acf}
\end{equation}
The sum runs over the $N$ pairs of points in the spectrum, indexed by $\nu$, which are not flagged as RFI or zeroed out for other reasons. In addition to correlating $\delta_{\rm on}$, the ACF is performed on several of the off-pulse spectra by defining $\bar{S}_{\rm smooth/off}(\nu)$ as the smoothed version of $S_{\rm off}(\nu)$ and $\delta_{\rm off}(\nu) = (S_{\rm off}(\nu) / \bar{S}_{\rm smooth/off}(\nu))-1$ for visual comparison.

\section{Scintillation Bandwidth Analysis}
 To measure the frequency dependence of $\nu_{dc}$, the ACF is computed in 40 MHz-wide subbands spanning the full $400-800$ MHz bandwidth of CHIME. The ACFs from each subband are shown in the left panel of Fig.~\ref{fig:acf} for an FRB which exhibits scintillation. To measure $\nu_{dc}$ for each subband, we fit to the data the following model as in~\citet{Masui_2015}:
\begin{equation} 
\mathbf{r}(\Delta \nu) = \dfrac{m}{\left(\dfrac{\nu_{dc}}{\SI{1}{\mega\hertz}}\right)^2 + \left(\dfrac{\Delta \nu}{\SI{1}{\mega\hertz}}\right)^2}.
\label{eq:cauchy}
\end{equation}
 The modulation index $m$ can be interpreted as the contrast between bright and dark scintles, where $m=1$ corresponds to complete constructive/destructive interference in bright/dark scintles, and $m=0$ corresponds to the absence of scintillation.
 If scintillation is the source of the observed frequency-domain correlations, the decorrelation bandwidth $\nu_{dc}$ is expected to follow a frequency-dependent power law:
\begin{equation} 
\nu_{dc}(\nu) \propto \left( \dfrac{\nu}{\SI{600}{\mega\hertz}}\right)^\alpha.
\label{eq:scaling}
\end{equation} 
The Milky Way exhibits density fluctuations consistent with the predictions of turbulence, in which $\alpha \approx 4$. The right panel of Fig. ~\ref{fig:acf} shows the power law scaling (Eqn.~\ref{eq:scaling}) of the $\nu_{dc}$ from each subband shown on the left. This value of $\alpha$ is also used to scale the NE2001 predictions for scintillation at 1 GHz to our central frequency (\SI{600}{\mega\hertz}).

\section{Results and Discussion}
\begin{figure*}[ht]
    \centering
            \includegraphics[width=0.31\textwidth]{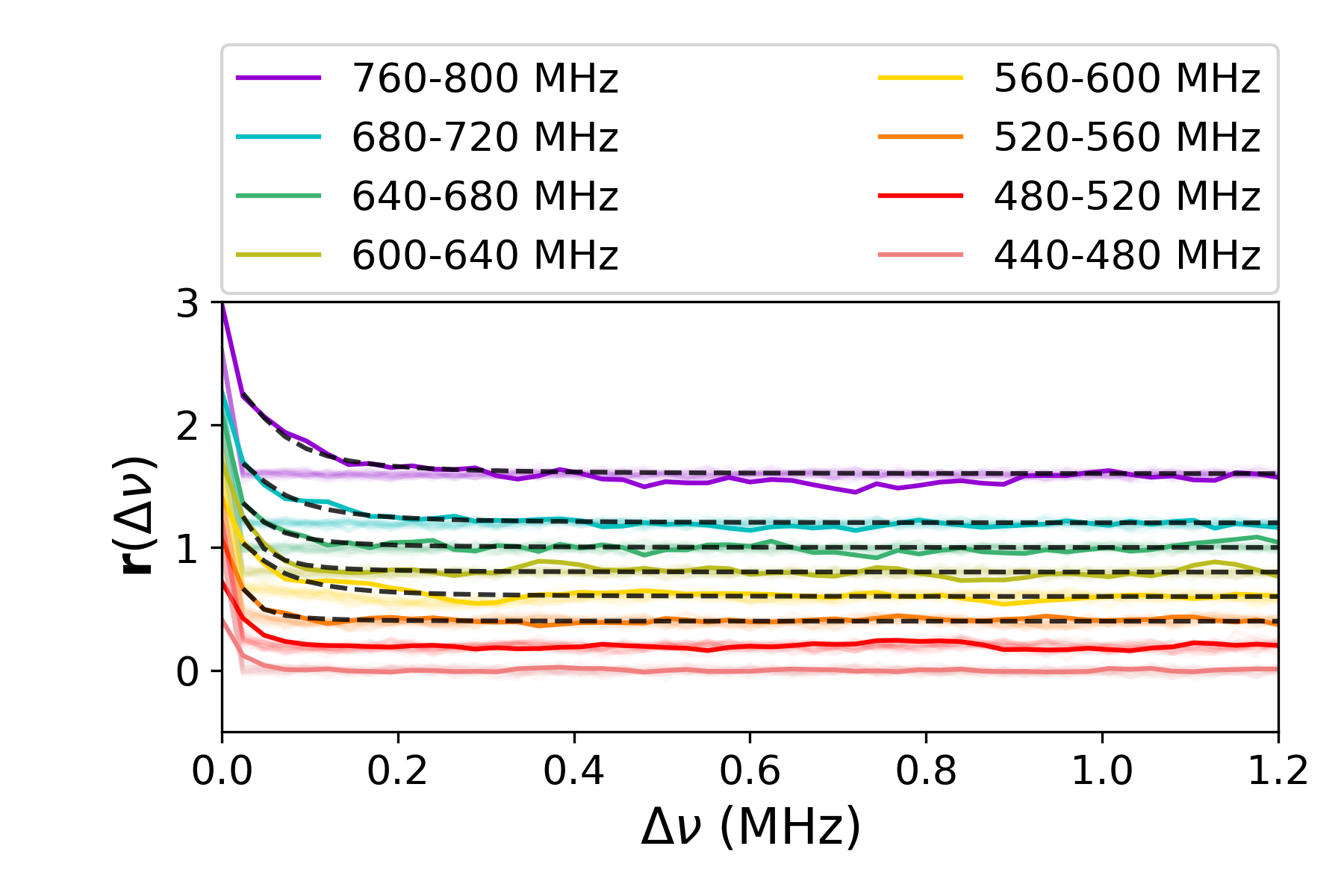} \includegraphics[width=0.31\textwidth]{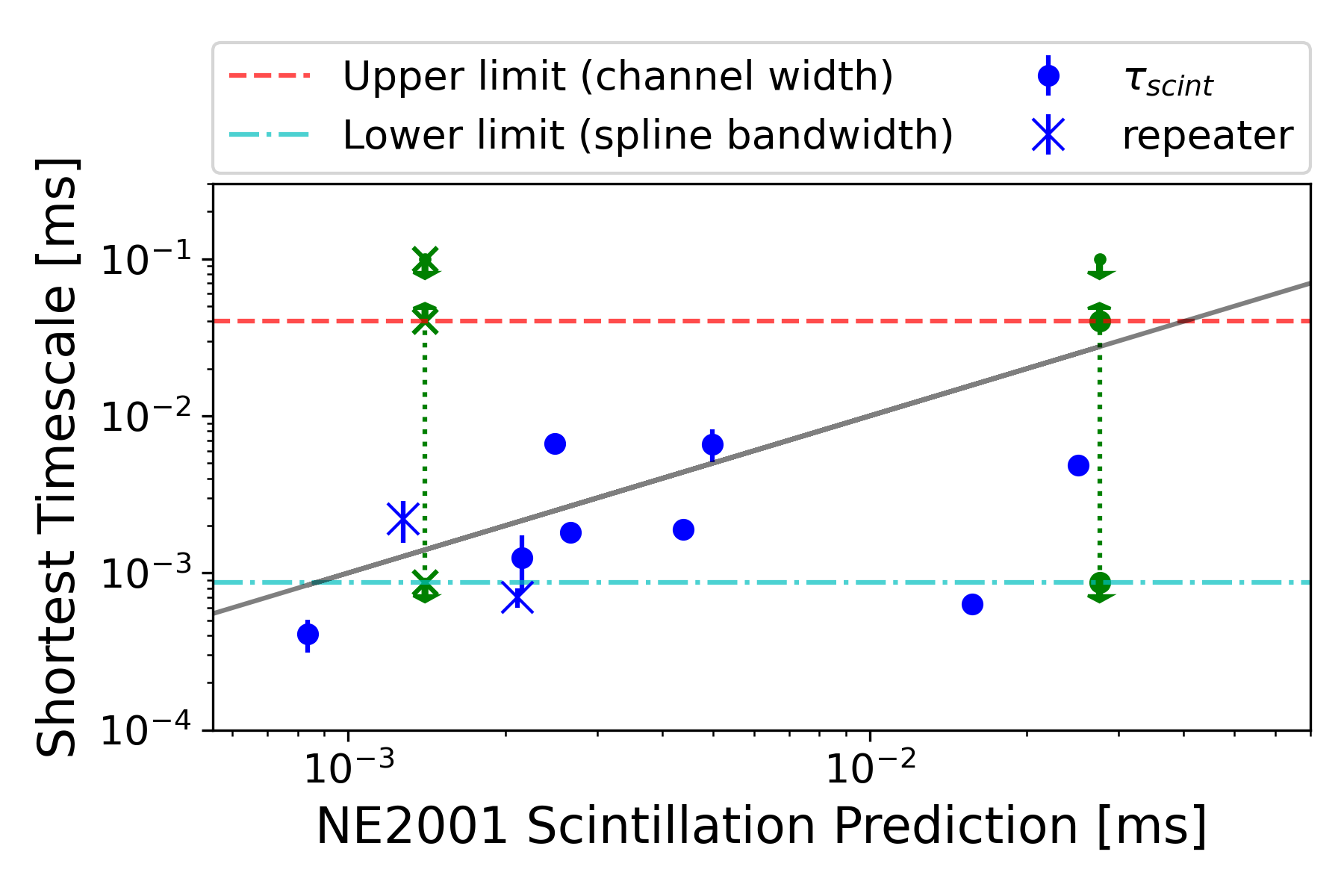}
            \includegraphics[width=0.31\textwidth]{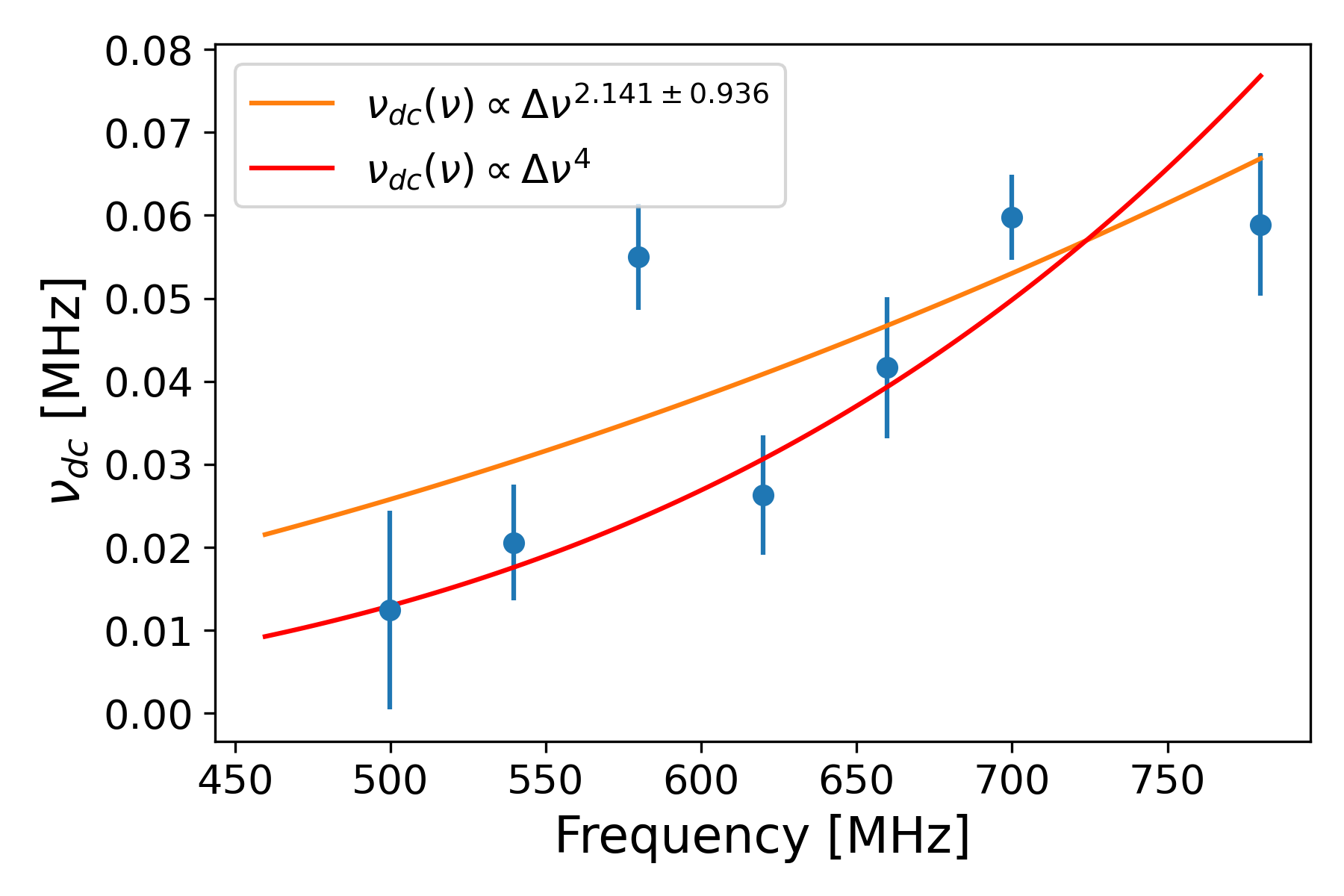}
            \caption{\textbf{Left:} The ACF of an FRB that exhibits scintillation. Subbands are offset by .2. \textbf{Right:} The frequency dependence of the decorrelation bandwidths. \textbf{Center:} The empirical distribution of scintillation timescales compared to NE2001 predictions. The gray line is the $y=x$ line. Green arrows represent upper/lower limits on scintillation based on non-measurement of scintillation in this analysis 
            and upper limits based on the non-observation of a scattering tail. This supports the interpretation of short scattering timescales in FRBs as originating from the Milky Way~\citep{Masui_2015}.}
    \label{fig:acf}
\end{figure*}

We have developed a pipeline to measure scintillation timescales for a sample of 12 FRBs using CHIME/FRB baseband data. For bursts without a measured scattering tail, our measurements constrain the total line-of-sight plasma inhomogeneities for that FRB.
In other cases where there is a pulse broadening timescale, the decorrelation bandwidth defines a second characteristic timescale. For these bursts, we plot the shortest timescale available, the observed scintillation timescale in the central panel of Fig.~\ref{fig:acf}, against the NE2001 prediction for the Galactic scintillation timescale, $\tau$, given by:
\begin{equation} 
\tau = C_1 / (2\pi \nu_{dc})
\label{eq:ne2001}
\end{equation}
where under the assumption of gaussian turbulence in the ISM $C_1 = 1$~\citep{cordes2002ne2001}. 
The near 1-to-1 correlation suggests that the Milky Way accounts for most of the scintillation observed in this sample of FRBs.

%

\vspace{5mm}






\bibliography{references}{}

\bibliographystyle{aasjournal}



\end{document}